\documentstyle[aps,prl,preprint]{revtex}
\begin{document}
\draft
\title{Evidence for Transition Temperature Fluctuation Induced Pinning in MgB$_2$ Superconductor}

\author{M. J. Qin, X. L. Wang, H. K. Liu, S. X. Dou}

\address{Institute for Superconducting and Electronic Materials, University of Wollongong, Wollongong, NSW 2522, Australia}

\maketitle

\begin{abstract}
The magnetic field dependent critical current density $j_c(B)$ of a MgB$_2$ bulk sample has been obtained by means of magnetization hysteresis measurements. The $j_c(B)$ curves at different temperatures demonstrate  a crossover from single vortex pinning to small-bundle vortex pinning, when the field is larger than the crossover field $B_{\rm sb}$.  The temperature dependence of the crossover field $B_{\rm sb}(T)$ is in agreement with a model of randomly distributed weak pinning centers via the spatial fluctuations of the transition temperature ($\delta T_c$-pinning), while pinning due to the mean free path fluctuations ($\delta l$-pinning) is not observed.

\end{abstract}
\vspace{1cm}

%\begin{multicols}{2}

The recent discovery of superconductivity in the intermetallic compound MgB$_2$ \cite{naga} with transition temperature at 39K has led to intensive experimental and theoretical activities \cite{larb,bugo,hink,bugos,jins,eomc,slus,canf,finn,budk,schm,taka,wenh,josh}, with the purpose of understanding the basic mechanism of superconductivity and the vortex pinning mechanism governing the critical current density $j_c$ in this new superconductor. Although the critical current density has been improved greatly since its discovery, the underlying pinning mechanism is still under investigation.

In type-II superconductors, the most important elementary interactions between vortices and pinning centers are the magnetic interaction and the core interaction \cite{grie,blat,word,thun,thune,beek,kesp}. The magnetic interaction arises from the interaction of surfaces between superconducting and non-superconducting material parallel to the applied magnetic field. In technical type-II superconductors with a high Ginzburg-Landau (GL) parameter $\kappa$, the magnetic interaction is usually very small and disappears with increasing magnetic field. The core interaction is usually more effective in technical type-II superconductors due to the short coherence length and the larger penetration depth (high $\kappa$). This interaction arises from the coupling of the locally distorted superconducting properties with the periodic variation of the superconducting order parameter. Two mechanisms of core pinning are predominant in type-II superconductors, i.e. $\delta T_c$- and $\delta l$-pinning. Whereas $\delta T_c$-pinning is caused by the spatial variation of the GL coefficient $\alpha$ associated with disorder in the transition temperature $T_c$, variations in the charge carrier mean free path $l$ near lattice defects are the main cause of $\delta l$-pinning.

It has been reported by Griessen et al. \cite{grie} that the $\delta l$-pinning mechanism is dominant in both YBa$_2$Cu$_3$O$_7$ and YBa$_2$Cu$_4$O$_8$ thin films. For the new superconductor MgB$_2$, a high $\kappa$ value of 26 has been reported \cite{finn}, it is therefore expected that the magnetic interaction is negligible, while the core interaction is more important. However, it has not been experimentally determined whether the $\delta l$-pinning or the $\delta T_c$-pinning is the dominant mechanism in MgB$_2$. The purpose of this Letter is to report measurements of the critical current density of this new material to achieve an understanding of the vortex pinning mechanism and to demonstrate that in MgB$_2$ governed by bulk pinning, $\delta T_c$ is the only important pinning mechanism.

All measurements have been performed on a MgB$_2$ bulk sample, which was prepared by conventional solid state reaction \cite{dous}. High purity Mg and B (amorphous) with a nominal composition ratio of Mg:B=1.2:2 were mixed and finely ground, then pressed into pellets 10 mm in diameter with 1-2 mm thickness. Extra Mg was added in order to make up for loss of Mg at high temperatures. These pellets were placed on an iron plate and covered with iron foil, then put into a tube furnace. The samples were sintered at temperatures between 700 and 1000$^{\mbox{o}}$C for 1-14h. A high purity Ar gas flow was maintained throughout the sintering process. A sample with $T_c=38.6$ K and dimensions of $2.18\times 2.76\times 1.88$ mm$^3$ was cut from the pellet. Phase purity was determined by XRD \cite{qinm} and grain size by SEM. Only a small level of MgO (less than 10\%) was found and the grain size was determined to be about 200 $\mu$m. 

Fig.\ \ref{fig1} shows the magnetization hysteresis loops of the MgB$_2$ sample every 2 K in the 14-36 K range. The results at lower temperatures, which have large flux jumping \cite{dous}, are not shown here. The symmetric magnetization hysteresis loops with respect to the magnetic field indicate the dominance of bulk current up to temperatures near $T_c$, rather than surface shielding current. Therefore, the bulk pinning is dominant in this sample, while the surface pinning is negligible. As a comparison, we show in the inset of Fig.\ \ref{fig1} the magnetization hysteresis loop of a pressed MgB$_2$ sample at 5 K. This sample is fabricated by pressing the MgB$_2$ powder into a pellet without sintering. The loop is highly asymmetric, showing a large reversible magnetization resulting from the surface current, and surface pinning plays an important role in this sample. No flux jumping is observed down to T=5 K, indicating that the grains in the sample are decoupled. The surface pinning effect has also been observed by Takano et al. \cite{taka} in their powder sample and bulk sample sintered at low temperature.

From these M(H) loops, we can calculate the critical current density using $j_c=\Delta M/a(1-a/3b)$, with $a$, $b$ the width and length of the sample perpendicular to the applied magnetic field, respectively. The resultant $j_c(B)$ curves at various temperatures are shown in a double logarithmic plot in Fig.\ \ref{fig2} as different symbols. As can be seen from the plateau at low magnetic field, $j_c$ initially has a weak dependence on the field. When the magnetic field is increased beyond a crossover field, it then begins to decrease quickly. The crossover field decreases with increasing temperature. Further increasing the magnetic field results in a faster drop in $j_c$ near the irreversibility line, which is obtained by using a criterion for the critical current density $j_c=100$ A/cm$^2$. The results are shown as open circles in Fig.\ \ref{irr}. The best fitting of the data yields the result,
\begin{equation}
B_{\rm irr}(t)=B_{\rm irr}(0)\left(1-t^2\right)^{\frac{3}{2}},
\label{irrline}
\end{equation}
shown as solid line in Fig.\ \ref{irr}, with $t=T/T_c$. For high temperature superconductors, a $(1-t)^{3/2}$ behavior is usually observed \cite{mull} and has been explained by means of giant flux creep \cite{yesh,tink}. The $(1-t)^{3/2}$ law is actually an approximation of the $(1-t^2)^{3/2}$ law as $t\to 1$, because $(1-t^2)\approx (1-t)(1+t)\approx 2(1-t)$. Also plotted in the figure is the $B_{c2}(T)$ data taken from Takano's work (the dashed line is just guide to the eye), showing that $B_{\rm irr}$ is well below $B_{c2}$. Therefore, giant flux creep also plays an important role in the new superconductor MgB$_2$. 

$j_c(B)$ characteristics very similar to those shown in Fig.\ \ref{fig2} have been observed by other groups \cite{bugo,wenh,josh}. Based on different physical assumptions on summation of the elementary pinning force $f_p$ to obtain the macroscopic pinning force $F_p$, different pinning models yield different $j_c(B)$ characteristics. The simplest model is the direct summation of $f_p$ to have $F_p=j_cB=n_pf_p$, where $n_p$ is the density of pinning centers in the sample. This strong pinning model neglects the influence of the flux line lattice. When the influence is taken into account, one have $F_p=j_cB=n_pf_p^2(u_0/f_p)d/a_0^2$, where $u_0$ is the maximum distortion of the flux line lattice, $d$ is the range of the pinning force, typically of the order of $xi$, and $a_0$ the flux line lattice constant. These two strong models yield $j_c(B)$ characteristics of $j_c\propto B^{-1}$ and $j_c\propto B^{-0.5}$, respectively. Due to the large densities of the pins $n_p$ and the small elementary interaction forces $f_p$, these two models are not representative for most real pinning systems \cite {bran}. For randomly distributed weak pinning centers, the macroscopic pinning force $F_p$ can be estimated using the basic concept of collective pinning \cite{lark}, which has been proved to be very successful in most real pinning systems,
\begin{equation}
F_p=j_cB=\sqrt{\frac{W}{V_c}}=\sqrt{\frac{W}{R_c^2L_c}}
\label{fp}
\end{equation}
with the correlation volume $V_c=R_c^2L_c$, with the correlation lengths $R_c$ perpendicular to the field direction and $L_c$ along the vortex line, and the pinning parameter $W=n_p<f_p^2>$. $R_c$ and $L_c$ depend on the applied magnetic field, the dimension of the flux line lattice (3D or 2D), and the elasticity or plasticity of the flux line lattice. For the 3D elastic flux line lattice, it has been derived by Blatter et al. \cite{blat} that $j_c$ is field-independent when the applied magnetic field is lower than the crossover field $B_{\rm sb}$ (single vortex pinning)
\begin{equation}
B_{\rm sb}=\beta_{\rm sb}\frac{j_{\rm sv}}{j_0}B_{c2}
\label{cross}
\end{equation}
where $\beta_{\rm sb}\approx 5$ is a constant, $j_0=4B_c/3\sqrt{6}\mu_0\lambda$ the depairing current, $B_c=\Phi_0/2\sqrt{2}\pi\lambda\xi$ the thermodynamic critical field, $B_{c2}=\mu_0\Phi_0/2\pi\xi^2$ the upper critical field and $j_{\rm sv}$ the critical current density in the single vortex pinning regime. When the applied field is larger than $B_{\rm sb}$ (small bundle pinning), $j_c(B)$ follows a exponential law 
\begin{equation}
j_c(B)\approx j_c(0)\exp\left[-\left(\frac{B}{B_0}\right)^{\frac{3}{2}}\right].
\label{explaw}
\end{equation}
When the applied magnetic field is larger than $B_{\rm lb}=\beta_{\rm lb}B_{c2}(j_{\rm sv}/j_0)[\ln(\kappa^2j_{\rm sv}/j_0)]^{2/3}$ (where $\beta_{\rm lb}$ is a constant $\approx 2$), this large bundle pinning regime is governed by a power law $j_c(B)\propto B^{-3}$. The 2D elastic flux line lattice shows single pancake pinning at low magnetic fields with field-independent $j_c$, then a 2D collective-pinning region at higher fields with $j_c\propto 1/B$. For fields higher than a crossover field $B_b^{\rm 3D}$, three-dimensional pinning is predicted.

From the characteristics of the $j_c(B)$ curves shown in Fig.\ \ref{fig2}, it is expected that the 3D elastic pinning model (single vortex pinning followed by small bundle pinning, then large bundle pinning) may  the dominant pinning mechanism in MgB$_2$. We therefore use Eq.(\ref{explaw}) to fit the $j_c(B)$ curves, with fitting parameters $j_c(0)$ and $B_0$. The fitting results for different temperatures are shown as solid lines in Fig.\ \ref{fig2}. At intermediate fields, Eq.(\ref{explaw}) fits the experimental data very well, while deviations from the fitting curves can be observed at both low and high fields. A clearer plot is shown in Fig.\ \ref{fig3}, where the $j_c(B)$ curve at 24 K is shown in a double-logarithmic plot of $-\log[j/j(B=0)]$ versus the applied field, which clearly shows a straight line at intermediate magnetic fields. The  deviation at low fields is denoted as $B_{\rm sb}$, indicating the crossover from the single vortex pinning regime to the small bundle pinning regime. The point of deviation at high fields was first considered as the crossover field from small bundle pinning to large bundle pinning. However, when we fit the $j_c(B)$ data at high fields to the power law $j_c(B)\propto B^{-n}$, $n$ is found to be as large as 20 rather than the theoretically predicted value of 3, making it unlikely that the system changes to the large bundle pinning regime. As the high field deviation is very close to the irreversibility line, which results from giant flux creep, it is likely that the deviation at high field may result from large thermal fluctuations, which lead to the rapid decrease in $j_c$, and therefore is denoted as $B_{\rm th}$, indicating thermal fluctuations. Other $j_c(B)$ expressions have also been tested for the fitting, such as $j_c(B)\propto 1/[1+(B/B_0)^n]$ and $j_c(B)\propto \exp[-(B/B_0)]$, but both yield poor fitting results. The inset of Fig.\ \ref{fig3} shows the $j_c(B)$ of the pressed sample at 5 K in a double-logarithmic plot, which indicates that when the surface pinning is important, the exponential drop in $j_c(B)$ [see Eq.(\ref{explaw})] no longer applies, but a power law $j_c(B)\propto B^{-1.2}$ is obvious.

The crossover field $B_{\rm sb}$ we have obtained between single vortex pinning and small bundle pinning as a function of temperature is shown in Fig.\ \ref{fig4} as open circles. We now compare the experimental data with theoretical predictions to get some insight into the pinning mechanism in MgB$_2$. Using $\lambda\propto (1-t^4)^{-1/2}$ and $\xi\propto [(1+t^2)/(1-t^2)]^{1/2}$, Griessen et al. have found that for $\delta l$-pinning the critical current density in the single vortex pinning regime $j_{\rm sv}\propto (1-t^2)^{5/2}(1+t^2)^{-1/2}$, while for $\delta T_c$-pinning $j_{\rm sv}\propto (1-t^2)^{7/6}(1+t^2)^{5/6}$. Inserting all these expressions into Eq.(\ref{cross}), we have 
\begin{equation}
B_{\rm sb}=B_{\rm sb}(0)\left(\frac{1-t^2}{1+t^2}\right)^{2/3}
\label{dtc}
\end{equation}
for $\delta T_c$-pinning, and 
\begin{equation}
B_{\rm sb}=B_{\rm sb}(0)\left(\frac{1-t^2}{1+t^2}\right)^2
\label{dl}
\end{equation}
for $\delta l$-pinning. The lines corresponding to Eqs.(\ref{dtc}) and (\ref{dl}) are indicated as $\delta T_c$-pinning and $\delta l$-pinning respectively in Fig.\ \ref{fig4}. The central result of this Letter is the remarkably good agreement found between $B_{\rm sb}$ and the corresponding $\delta T_c$-pinning line in the figure. In sharp contrast, the $\delta l$-pinning line shown in Fig.\ \ref{fig4} is in total disagreement with the experimental data.

Having derived the crossover fields $B_{\rm sb}$ and $B_{\rm th}$, we now reconstruct the B-T phase diagram shown in Fig.\ \ref{irr}. The final B-T phase diagram is shown in Fig.\ \ref{bt}. The vortex solid region is divided into three smaller regions. Dingle vortex pinning governs the region below $B_{\rm sb}$, between $B_{\rm sb}$ and $B_{\rm th}$, small bundle pinning becomes dominant, while between $B_{\rm th}$ and $B_{\rm irr}$, thermal fluctuations are more important. Large flux bundle pinning is not observed in MgB$_2$, but may be concealed by the thermal fluctuation effects.

In summary, we have found strong evidence for $\delta T_c$-pinning, i.e. pinning via the spatial fluctuations in the transition temperature, in the new superconductor MgB$_2$, while $\delta l$-pinning i.e., pinning via the spatial fluctuations of the charge carrier mean free path, is not observed. The B-T phase diagram of the MgB$_2$ sample has been derived, showing that at low fields below $B_{\rm sb}$ the system is dominated by single vortex pinning and changes to smaller bundle pinning when $B>B_{\rm sb}$. When $B>B_{\rm th}$, this region in the vortex solid area is dominated by thermal fluctuations. The irreversibility line may result from the giant flux creep effect.

The authors would like to thank the Australian Research Council for financial support.

\begin{figure}
\caption{Magnetization hysteresis loops of the MgB$_2$ bulk sample taken every 2 K in the 14-36 K range. Results at lower temperatures are not shown because of large flux jumping. Inset shows the magnetization hysteresis loop of a pressed MgB$_2$ sample at 5 K, showing large reversible magnetization from the surface current.}
\label{fig1}
\end{figure}

\begin{figure}
\caption{Critical current density $j_c$ calculated from the Bean critical state model, indicated by different symbols for different temperatures. Solid lines are fitting curves using Eq.(\ref{explaw}).}
\label{fig2}
\end{figure}

\begin{figure}
\caption{B-T phase diagram of the new superconductor MgB$_2$. Open circles represent the irreversibility line obtained from Fig.\ \ref{fig2}, and the solid line is a fit to $B_{\rm irr}=5.2(1-t^2)^{3/2}$. Solid circles represent the $B_{c2}(T)$ line from Takano's data using resistive measurements. The dashed line is just a guide to the eye.}
\label{irr}
\end{figure}

\begin{figure}
\caption{Critical current density at 24 K in a double-logarithmic plot of $-\log[j/j(B=0)]$ versus the applied field. The solid line is a fit using Eq.(\ref{explaw}). $B_{\rm sb}$ indicates the crossover field from single vortex pinning to small bundle pinning, while $B_{\rm th}$ is the crossover field to the thermal fluctuations dominated regime. Inset shows the $j_c(B)$ curve of the pressed MgB$_2$ sample with a large contribution from the surface current, which shows a $B^{-1.2}$ behavior.}
\label{fig3}
\end{figure}

\begin{figure}
\caption{Temperature dependence of the crossover field $B_{\rm sb}$. The $\delta T_c$-pinning line corresponds to Eq.(\ref{dtc}), which is in agreement with the experimental data, while The $\delta l$-pinning line corresponds to Eq.(\ref{dl}), which is not in agreement with the experimental data.}
\label{fig4}
\end{figure}

\begin{figure}
\caption{B-T phase diagram of the new superconductor MgB$_2$. $B_{\rm sb}(T)$ is the fitting curve of Eq.(\ref{cross}) to the experimental data (see Fig.\ \ref{fig4}). $B_{\rm irr}(T)$ is the fitting curve of Eq.(\ref{irrline}) to the experimental data (see Fig.\ \ref{irr}). $B_{\rm th}$ is the crossover field to thermal dominant region (see Fig.\ \ref{fig4}). Again the $B_{c2}(T)$ line is taken from Takano's data using resistive measurements.}
\label{bt}
\end{figure}

%\end{multicols}
\end{document}